\documentclass[journal=nalefd,manuscript=letter]{achemso}
\usepackage{xfrac}
\usepackage{amssymb}
\usepackage{braket}
\usepackage{graphicx}
\usepackage{verbatim}
\usepackage{etoolbox}
\usepackage{amsfonts}
\usepackage{amsmath}
\usepackage[utf8]{inputenc}
\usepackage{mathtools}
\usepackage{soul,xcolor,colortbl}
\usepackage{hyperref} \hypersetup{colorlinks=true,citecolor=blue,linkcolor=blue,urlcolor=blue}
\usepackage{tabularx}
\usepackage{lscape}

\usepackage{multirow}
\usepackage{longtable}
\usepackage{cellspace}
\usepackage{resizegather}
\usepackage{dcolumn}
\usepackage[normalem]{ulem}
\usepackage{pdfpages}

\usepackage{bm}
\usepackage{array}
\newcolumntype{C}[1]{>{\centering\arraybackslash}p{#1}}\usepackage{soul}
\definecolor{Gray}{gray}{0.85}

\usepackage{color}
\usepackage{morefloats}

\definecolor{Gray}{gray}{0.9}
\definecolor{LightCyan}{rgb}{0.88,1,1}
\definecolor{green}{rgb}{0.5451,0.2706,0.0745}

\def\DFT{{\small  DFT}}
\def\CCE{{\small  CCE}}
\def\LDA{{\small  LDA}}

\def\PBE{{\small  PBE}}
\def\SCAN{{\small  SCAN}}

\def\FM{{\small FM}}
\def\AFM{{\small AFM}}

\def\AFLOW{{\small AFLOW}}
\def\VASP{{\small VASP}}

\title{A New Group of Two-Dimensional Non-van der Waals Materials with Ultra Low
Exfoliation Energies}

\author{Tom Barnowsky}
\affiliation{Institute of Ion Beam Physics and Materials Research, Helmholtz-Zentrum Dresden-Rossendorf, 01328 Dresden, Germany}
\alsoaffiliation{Theoretical Chemistry, Technische Universit\"{a}t Dresden, 01062 Dresden, Germany}
\author{Arkady V. Krasheninnikov}
\affiliation{Institute of Ion Beam Physics and Materials Research, Helmholtz-Zentrum Dresden-Rossendorf, 01328 Dresden, Germany}
\alsoaffiliation{Department of Applied Physics, Aalto University, Aalto 00076, Finland}
\author{Rico Friedrich}
\email{r.friedrich@hzdr.de}
\affiliation{Institute of Ion Beam Physics and Materials Research, Helmholtz-Zentrum Dresden-Rossendorf, 01328 Dresden, Germany}
\alsoaffiliation{Theoretical Chemistry, Technische Universit\"{a}t Dresden, 01062 Dresden, Germany}

\date{\today}

\begin{document}

\maketitle

\begin{abstract}

The exfoliation energy --- quantifying the energy required to extract a two-dimensional (2D) sheet from the surface of a bulk material --- is a key parameter determining the synthesizability of 2D compounds.
Here, using \emph{ab initio} calculations, we present a new group of non-van der Waals 2D materials derived from non-layered crystals which exhibit ultra low exfoliation energies.
In particular for sulfides, surface relaxations are essential to correctly describe the associated energy gain needed to obtain reliable results.
Taking into account long-range dispersive interactions has only a minor effect on the energetics and ultimately proves that the exfoliation energies are close to the ones of traditional van der Waals bound 2D compounds.
The candidates with the lowest energies, 2D SbTlO$_3$ and MnNaCl$_3$, exhibit appealing electronic, potential topological, and magnetic features as evident from the calculated band structures making these systems an attractive platform for fundamental and applied nanoscience.

\noindent

\vspace{0.2cm}

\end{abstract}

\noindent
The discovery of new two-dimensional (2D) materials --- traditionally derived from bulk layered compounds held together by weak van der Waals (vdW) forces ---
outlined over the last two decades a diverse zoo of representatives
showcasing  unique topological \cite{Kou_JPCL_2017}, electronic \cite{Butler_ACSNano_2013,Chhowalla_NChem_2013,Manzeli_NatRevMat_2017}, magnetic \cite{Burch_Nature_2018,Gibertini_NNano_2019,Huang_SmartMat_2021}, and superconducting \cite{Cao_Nature_2018,Campi_NanoLett_2021} properties.
The weak interaction between the structural units in their bulk counterparts leads to a natural geometric separation of the
2D subunits in the crystals, giving rise to the possibility of mechanical \cite{Novoselov_PNAS_2005} and
liquid-phase \cite{Nicolosi_Science_2013} exfoliation.
This class of nanostructures thus opens up prospects for fundamental research in reduced dimensions as well as for various applications in the energy sector \cite{Anasori_NatRevMat_2017,Wang_NCom_2021,Chepkasov_NanoEn_2020}, catalysis \cite{Deng_NNano_2016,Wang_ChemRev_2019}, and opto-electronics \cite{Wang_NNano_2012,Butler_ACSNano_2013,Lemme_MRSBul_2014}.
The large scale deployment of these vdW 2D materials in modern technologies is, however, still very limited \cite{Lemme_NCOM_2022}.

In light of the ubiquitous use of many standard non-layered materials in research and technology for which handling and processing is well established, the search for non-vdW 2D materials is appealing \cite{Balan_MatTod_2022}.
Recently and somewhat unexpectedly, atomically thin 2D sheets derived from non-vdW bonded oxides were indeed manufactured.
The first representatives realized in experiment by a special chemical exfoliation process were hematene \cite{Puthirath_Balan_NNANO_2018} and ilmenene \cite{Puthirath_Balan_CoM_2018} obtained from the earth-abundant ores hematite ($\alpha$-Fe$_2$O$_3$) and  ilmenite (FeTiO$_3$), followed by a few others \cite{Yadav_AdvMatInt_2018,Puthirath_Small_2020,Kaur_ACSNano_2020,Puthirath_JPCC_2021,Moinuddin_AdvElMat_2021,Hu_two-dimensional_2021,Yousaf_JPCC_2021,Guo_Nanoscale_2021,Gibaja_AdvMat_2021,Xu_ChemSci_2021,Peng_NChem_2021,Homkar_ACSApplNanoMat_2021,Chen_FlatChem_2022,Toksumakov_arxiv_2022}.
Unlike, \emph{e.g.} silicene or borophene \cite{Mannix_NatNano_2018}, these materials do not need to strongly interact with a substrate to be stable.
From the computational side, a recent data-driven search on non-vdW 2D systems outlined 28 candidates with a variety of appealing electronic and magnetic properties \cite{Friedrich_NanoLett_2022}.
First application-oriented studies indicate promising perspectives
for opto-electronics \cite{Wei_CMS_2020},
photo-catalytic activity for water splitting \cite{Puthirath_Balan_NNANO_2018,Puthirath_Balan_CoM_2018}, and photoconductivity \cite{Hu_two-dimensional_2021}.
Despite these early successes, the non-vdW 2D materials space is still very narrow and it remains to be understood what promotes the exfoliability of non-vdW bulk systems into 2D sheets.

A key quantity determining the synthesizability of 2D materials is the exfoliation energy $\Delta E_{\mathrm{exf}}$.
It represents the energy needed to peel off one 2D sheet from the surface of the bulk parent material.
It can be computed accurately according to the method of Jung \emph{et al}. \cite{Jung_NanoLett_2018} at minimal computational cost, who proved that the exfoliation energy is the energy difference between an isolated 2D sheet and one such facet in the bulk, also known as inter-layer binding energy.
For the standard reference material graphene, the exfoliation energy is known to be $\sim$20~meV/\AA$^2$ from both experiment and advanced density functional theory (\DFT) \cite{Zacharia_PRB_2004,Bjorkman_PRL_2012}.
It was even found that this value appears to be universal for many layered systems largely independent of their electronic structure.
However, later on a rather complete screening of the vdW 2D materials space outlined a much wider distribution of exfoliation energies over several ten meV/\AA$^2$ \cite{Mounet_AiiDA2D_NNano_2018}, likely due to contributions from other interactions than just vdW.
Based on detailed energetic and structural considerations, upper bounds of $\sim$130~meV/\AA$^2$ and 200~meV/atom for the exfoliation energy have been proposed to consider a material as (classically) exfoliable \cite{Choudhary_SciRep_2017,Mounet_AiiDA2D_NNano_2018}.
As such, the calculated value of $\sim$140~meV/\AA$^2$ ($\sim$310~meV/atom) for hematene \cite{Wei_JPCC_2020,Friedrich_NanoLett_2022} seems surprisingly high, yet successful liquid-phase isolation of the 2D sheet was reported \cite{Puthirath_Balan_NNANO_2018}.

In the recent data-driven search for non-vdW 2D systems, it was indicated that for several of the 28 oxidic candidates, the exfoliation energy is significantly smaller.
For a few systems, it even comes close to the one of graphene if the surface cations are in a low ($+1$) oxidation state \cite{Friedrich_NanoLett_2022} suggesting that mechanical peel-off might be feasible.
This effect was rationalized through the minimization of electrostatic interactions between the 2D sheets in case of small surface charges.
While the consideration of the systems in vacuum is an idealization with respect to experiment, it allows for an assessment of the fundamental exfoliation energetics of these materials.
Here, we focus explicitly on novel candidates with surface cations in low oxidation states
and generalize the previous efforts by investigating also non-oxides.
The goal is to try to find an answer to the question:
How small can the exfoliation energy of a non-vdW 2D material get?

\begin{figure*}[ht!]
	\centering
	\includegraphics[width=\textwidth]{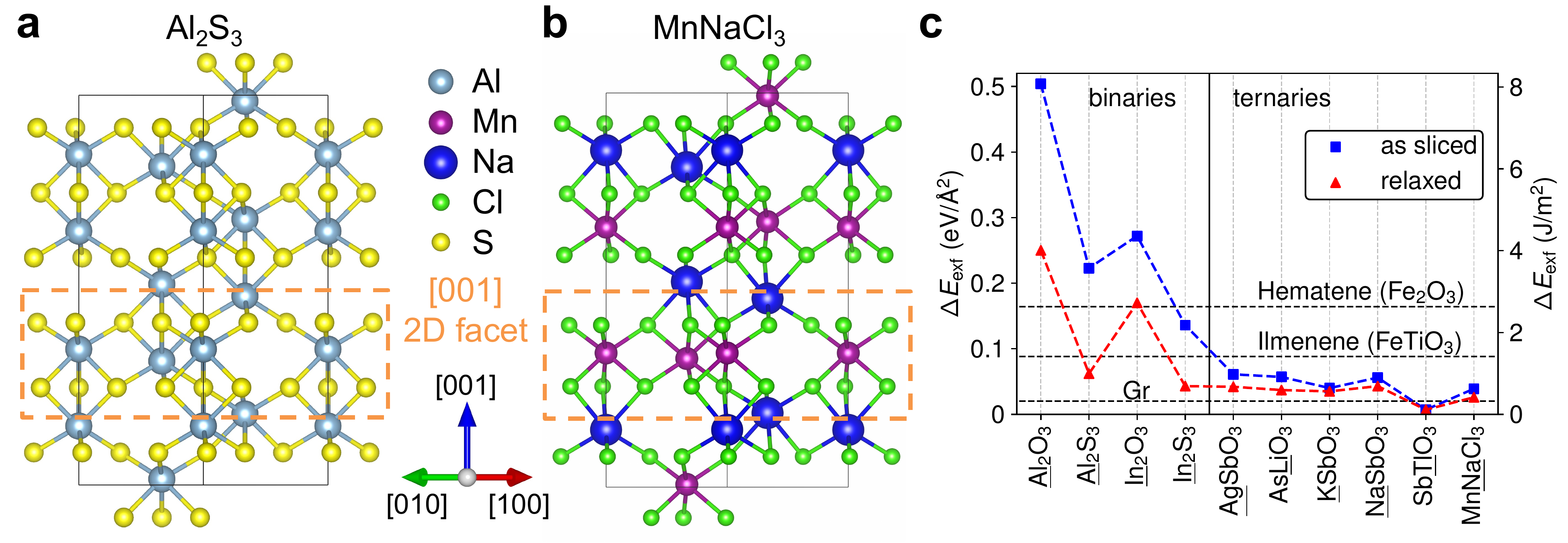}
	\caption{\small \textbf{Structures and exfoliation energies.}
	Atomic structure of (\textbf{a}) Al$_2$S$_3$ and (\textbf{b}) MnNaCl$_3$\cite{VESTA}.
	The exfoliable [001] facet (monolayer) is indicated in the orange dashed box.
	The black line denotes the conventional unit cell.
	The compass indicating crystal directions applies to both figures.
	(\textbf{c}) Calculated exfoliation energies from \SCAN\ without (as sliced) and with (relaxed) structural optimization of the 2D sheets.
	As a reference, the exfoliation energy of graphene (Gr) \cite{Zacharia_PRB_2004,Mounet_AiiDA2D_NNano_2018} as well as for hematene and ilmenene \cite{Friedrich_NanoLett_2022} are indicated by the dashed horizontal black lines.
	For the ternaries, the data for the slabs with the energetically favorable termination are plotted and for all systems, the terminating element is underlined at the bottom axis.
	The dashed lines connecting the data points are visual guides.
    }
	\label{fig1}
\end{figure*}

The input structures are retrieved via the \AFLOW\ APIs \cite{curtarolo:art92,curtarolo:art128,Oses_arxiv_2022} and web interfaces \cite{curtarolo:art75,Esters_arxiv_2022} as well as the library of crystallographic prototypes \cite{curtarolo:art121}.
The structures of Al$_2$S$_3$ and MnNaCl$_3$ are depicted in Figs~\ref{fig1}(\textbf{a}) and (\textbf{b}) as examples to visualize the structural prototype of the investigated systems.
The exfoliated [001] facets are also indicated.
The dynamic stability of the outlined 2D systems is verified as follows:
First a 2$\times$2 in-plane supercell is constructed from the relaxed structures.
Then, the atomic coordinates are randomized (gaussian distribution, standard deviation 50~m\AA) \cite{ase,Larsen_JPhysCM_2017} and the structures are reoptimized.
In all cases, the slabs relax back to the previous geometry.
Although they also exhibit the same structure, Yb$_2$S$_3$ and Lu$_2$S$_3$ are not considered here as the corresponding 2D sheets were found to be dynamically unstable during the test.

\noindent
\textbf{{\emph{Ab initio} Exfoliation Energies}.}
In Fig.~\ref{fig1}(\textbf{c}), the calculated exfoliation energies are depicted also including, for comparison to the considered sulfides, the previously obtained values for the corresponding 2D oxides Al$_2$O$_3$ and In$_2$O$_3$ \cite{Friedrich_NanoLett_2022}.
For the ternaries, the sheets can be terminated by either of the two cation species.
Here, only the results for the energetically favorable termination are plotted and a comparison to the unpreferred termination is presented in Fig.~S1 in the Supporting Information ({\small SI}).
The $\Delta E_{\mathrm{exf}}$ of the newly considered systems extend over a large range of more than an order of magnitude.
Ultra low values are achieved for the systems with surface cations in $+1$ oxidation states (such as Ag$^+$, Li$^+$, K$^+$, Na$^+$, and Tl$^+$) being as small as 7~meV/\AA$^2$ for SbTlO$_3$
and 26~meV/\AA$^2$ for MnNaCl$_3$.
These numbers are comparable to or even below the graphene reference value of $\sim$20~meV/\AA$^2$.
They are significantly smaller than for any other non-vdW 2D system considered before.

The blue curve denotes the results obtained when omitting structural relaxations of the extracted 2D facets, \emph{i.e.} keeping them ``as sliced'' from the bulk, as is common reliable practice for traditional vdW 2D systems \cite{Mounet_AiiDA2D_NNano_2018}.
These values reduce significantly for almost all candidates when including structural optimization as denoted by the red curve pinpointing at the crucial role of relaxations for the energetics of these materials.
For the binary sulfides, this effect is particularly pronounced.
Not only is the absolute value of the ``relaxed'' exfoliation energy of the sulfides a factor of four lower than for the corresponding oxides (62~meV/\AA$^2$ for Al$_2$S$_3$ \emph{vs.} 250~meV/\AA$^2$ for Al$_2$O$_3$ and 43~meV/\AA$^2$ for In$_2$S$_3$ \emph{vs.} 170~meV/\AA$^2$ for In$_2$O$_3$), but also the relative change from ``as sliced'' to ``relaxed'' is much stronger:
While for the oxides the exfoliation energy reduces by about a factor of two upon relaxation (from 504~meV/\AA$^2$ to 250~meV/\AA$^2$ for Al$_2$O$_3$ and from 272~meV/\AA$^2$ to 170~meV/\AA$^2$ for In$_2$O$_3$), for the sulfides, it decreases by about a factor of three to four (from 223~meV/\AA$^2$ to 62~meV/\AA$^2$ for Al$_2$S$_3$ and from 136~meV/\AA$^2$ to 43~meV/\AA$^2$ for In$_2$O$_3$).
For the other systems, the change in the values based on structural relaxations is less pronounced and amounts to maximally a factor of $\sim$1.5 (for AsLiO$_3$ the reduction is from 57~meV/\AA$^2$ to 37~meV/\AA$^2$).

\begin{figure*}[ht!]
	\centering
	\includegraphics[width=\textwidth]{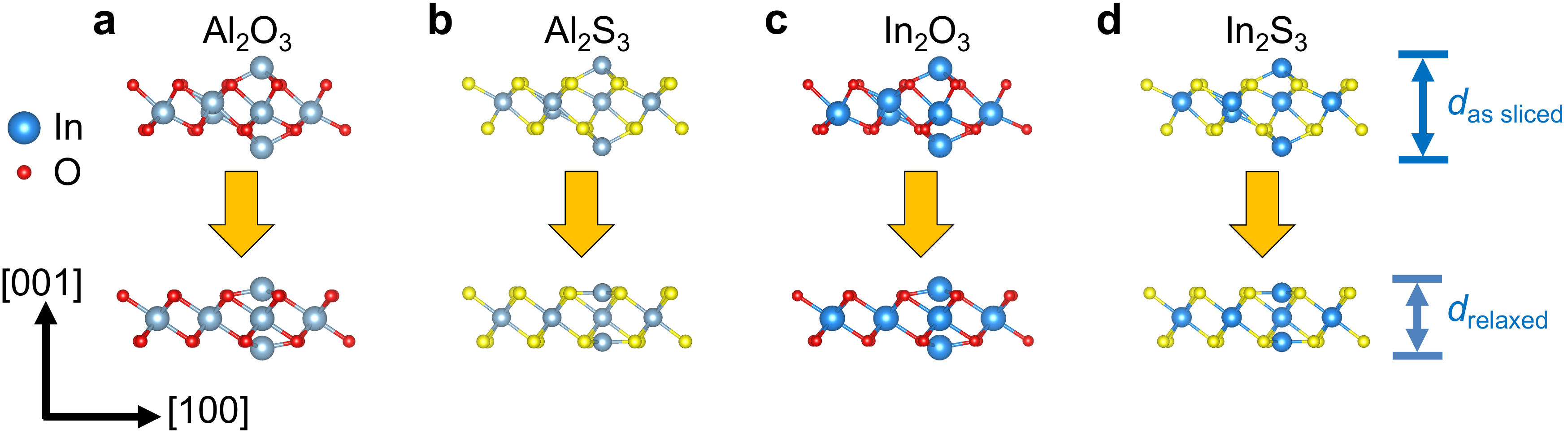}
	\caption{\small \textbf{Structural Relaxations.}
	Side view of the unit cells of ``as sliced'' (top) and ``relaxed'' (bottom) 2D (\textbf{a}) Al$_2$O$_3$, (\textbf{b}) Al$_2$S$_3$, (\textbf{c}) In$_2$O$_3$, and (\textbf{d}) In$_2$S$_3$.
    }
	\label{fig2}
\end{figure*}

\noindent
\textbf{{The Role of Structural Relaxations}.}
The origin of this strong change in energy in case of the sulfides as compared to the oxides can be traced back to intricate structural changes.
The side views in Fig.~\ref{fig2} indicate that the 2D systems lower their energy by an inward relaxation of the terminating cations leading to an overall (vertical) thickness reduction from $d_{\mathrm{as\ sliced}}$ to $d_{\mathrm{relaxed}}$.
This effect is more pronounced in case of the sulfides as compared to the oxides.
While Al$_2$O$_3$ (In$_2$O$_3$) relaxes from a thickness of 3.84~\AA\ (4.14~\AA) to 2.72~\AA\ (3.06~\AA), Al$_2$S$_3$ (In$_2$S$_3$) goes from 4.94~\AA\ (5.16~\AA) to 2.96~\AA\ (3.23~\AA).
Hence, although the absolute thickness of the sulfides is larger, the absolute \emph{reduction} by $\sim$2~\AA\ is about twice as large as for the oxides ($\sim$~1.1~\AA).
Thus, the stronger structural relaxation due to the larger anion allows the surface cation to dive deeper into the anion plane leading to an almost planar coordination of the surface cations in the sulfides.
Thereby, the energy is lowered more efficiently leading to stronger reductions in $\Delta E_{\mathrm{exf}}$.

As a response to the vertical contraction, the oxides stretch laterally, \emph{i.e.} the in-plane lattice constant of Al$_2$O$_3$ (In$_2$O$_3$) increases from 4.75~\AA\ (5.51~\AA) to 4.86~\AA\ (5.62~\AA).
For the sulfides, this change is less pronounced as for Al$_2$S$_3$ (In$_2$S$_3$) the value goes from 6.03~\AA\ (6.59~\AA) to 6.07~\AA\ (6.59~\AA).

\noindent
\textbf{{Contribution of Long-range vdW Interactions}.}
\begin{figure*}[t!]
	\centering
	\includegraphics[width=\textwidth]{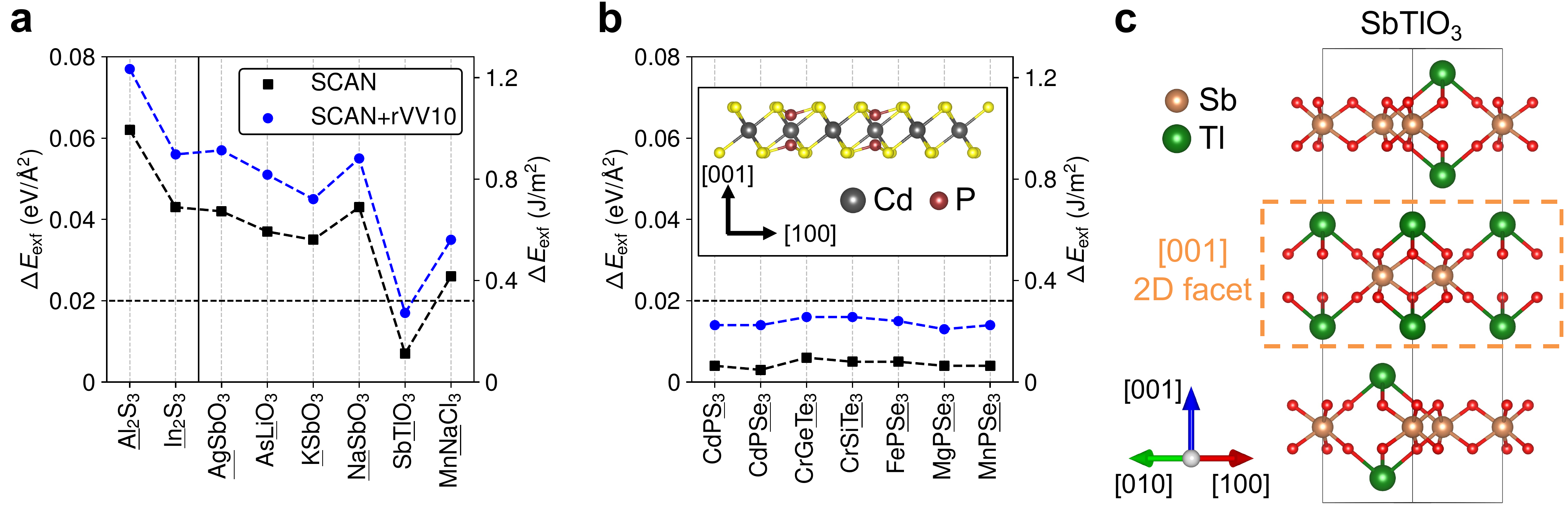}
	\caption{\small \textbf{Contribution of long-range vdW interactions.}
	Comparison of the calculated exfoliation energies from \SCAN\ and \SCAN+r{\small VV10} for (\textbf{a}) the eight non-vdW 2D candidates and (\textbf{b}) seven vdW 2D systems with the same structure.
	The dashed horizontal black line indicates the graphene reference value\cite{Zacharia_PRB_2004,Mounet_AiiDA2D_NNano_2018}.
	The dashed lines connecting the data points are visual guides.
	Inset in (\textbf{b}): side view of CdPS$_3$.
	The vertical black line in (\textbf{a}) separates binaries from ternaries.
	(\textbf{c}) Atomic structure of SbTlO$_3$.
	The exfoliable [001] facet (monolayer) is indicated in the orange dashed box.
	The black line denotes the conventional unit cell.
    }
	\label{fig3}
\end{figure*}
When exfoliation energies get as small as for graphene or even lower, the question of the importance of long-range vdW interactions naturally arises.
While the employed \SCAN\ functional has been pointed out to capture intermediate-range vdW interactions relevant for \emph{e.g.} hydrogen bonds \cite{Sun_NChem_2016}, it does not include long-range dispersion contributions.
These can be accounted for within the \SCAN+r{\small VV10} scheme\cite{Peng_PRX_2016}.
We have thus also calculated the exfoliation energies for all systems with this approach and compare them to the plane \SCAN\ results in Fig.~\ref{fig3}(\textbf{a}).
As expected, inclusion of the long-range interactions increases the exfoliation energies but only by a rather constant shift of 10-15~meV/\AA$^2$ --- the typical order of magnitude for dispersive interactions.
Hence no qualitative change in the exfoliation behavior is anticipated from this.

As a further comparison, there are seven materials (CdPS$_3$, CdPSe$_3$, CrGeTe$_3$, CrSiTe$_3$, FePSe$_3$, MgPSe$_3$, and MnPSe$_3$) in the \AFLOW\ database with the same structural prototype
that are clearly layered, \emph{i.e.} traditional vdW bonded materials as evident from visual inspection of the bulk geometries.
Thus, the same structural prototype can host both vdW and non-vdW bonded 2D materials.
The exfoliation energies for these well established 2D systems were also computed with the same methods and are provided in Fig.~\ref{fig3}(\textbf{b}).
Here, dispersive interactions are essential to obtain reliable absolute values as the constant shift by $\sim$10~meV/\AA$^2$ increases the \SCAN\ results by a factor of three.
For the non-vdW systems in Fig.~\ref{fig3}(\textbf{a}), the \SCAN\ values account already for 70-80\% of the \SCAN+r{\small VV10} result
since here the largest contribution to the bonding does not derive from long-range interactions.
This consideration of the different interaction contributions therefore provides a clear distinction between non-vdW and vdW 2D materials.

SbTlO$_3$ is a special case at the intersection between the vdW and non-vdW 2D materials spaces.
The optimized bulk (layered) structure with the exfoliable facet highlighted is depicted in Fig.~\ref{fig3}(\textbf{c}).
It is structurally equivalent to the other non-vdW systems as the Tl cations are at the surface of the 2D sheets in contrast to traditional 2D systems such as the ones from Fig.~\ref{fig3}(\textbf{b}) which are terminated by the anions (see the side view of CdPS$_3$ in the inset as an example).
Yet, the exfoliation energy of SbTlO$_3$ appears to be largely governed by long-range vdW contributions (see Fig.~\ref{fig3}(\textbf{a})).
However, the intermediate range vdW interaction captured by the SCAN exfoliation energy accounts with $\sim$40\% for a larger portion of the total exfoliation energy (from \SCAN+r{\small VV10}) compared to the traditional vdW 2D compounds of Fig.~\ref{fig3}(\textbf{b}) where this amounts on average to only $\sim$30\%.
As a result of these characteristics, this compound was already identified as a potential 2D material by Mounet \emph{et al.} \cite{Mounet_AiiDA2D_NNano_2018} although not discussed in detail.

An important general remark regarding the exfoliation energies must be made.
While it is well known that the standard \DFT\ functional \PBE($+U$) tends to underestimate binding energies (overestimating bond lengths also known as underbinding, see also the comparison including the \PBE$+U$ exfoliation energies in Fig.~{\small SI}2), there are several indications that the employed \SCAN\ functional tends to overestimate covalent and ionic binding energy contributions (for oxides).
Firstly, it has been shown that the \SCAN\ results are usually very close to the exfoliation energies computed from \LDA\ \cite{Friedrich_NanoLett_2022} for which overbinding effects are well established.
Secondly, it has been demonstrated that \SCAN\ systematically overestimates oxide formation enthalpies similar to \LDA\ \cite{Friedrich_CCE_2019}, again indicating binding effects to be on the high side.
\SCAN$+$rVV10 was shown to compute long-range vdW contributions reliably with no particular bias with respect to standard reference results from the random phase approximation \cite{Peng_PRX_2016,Emrem_AdvThSim_2022}.
Thus, we expect the \SCAN$+$rVV10 results to provide an upper bound for the exfoliation energies of non-vdW 2D materials.

\noindent
\textbf{{Band Structures and Magnetic Properties of SbTlO$_3$ and MnNaCl$_3$.}}
\begin{figure*}[ht!]
	\centering
	\includegraphics[width=\textwidth]{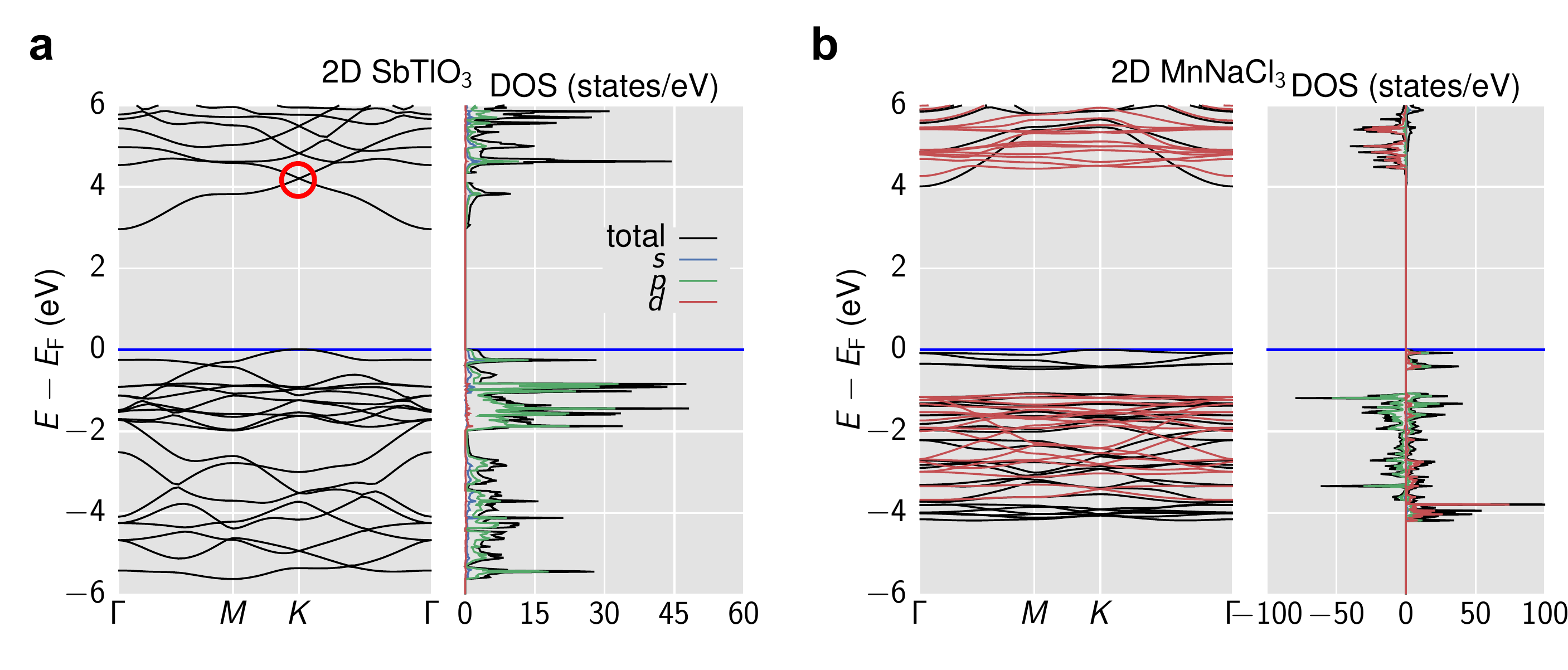}
	\caption{\small \textbf{Band structures and densities of states.}
	Band structure and density of states of (\textbf{a}) 2D SbTlO$_3$ and (\textbf{b}) 2D MnNaCl$_3$.
	The energies are aligned at the respective Fermi energy $E_{\mathrm{F}}$.
	For 2D SbTlO$_3$, a linear band crossing at the $K$-point is highlighted by the circle.
	For the spin polarized band structure in (\textbf{b}), majority spin bands (positive DOS) are indicated in black while minority spin
bands (negative DOS) are in red.
    }
	\label{fig4}
\end{figure*}
To showcase the potential of these materials, we briefly discuss the band structure and magnetic properties of SbTlO$_3$ and MnNaCl$_3$ --- the two systems with the lowest exfoliation energies --- and present the results for all compounds in comparison to the respective bulk bands in the {\small SI}.

According to the band structures in Fig.~\ref{fig4}, both systems are large gap insulators (calculated band gaps: 2.94~eV and 4.01~eV) with valence band maximum and conduction band minimum at the $K$- and $\Gamma$-points, respectively.
An appealing feature for SbTlO$_3$ is the Dirac cone like linear band crossing at the high-symmetry $K$-point at about 4.2~eV above the Fermi level which might be readily accessible via moderate doping.
This feature might hint at interesting topological properties calling for further investigations for instance addressing the explicit calculation of topological invariants.
MnNaCl$_3$ on the other hand shows ferromagnetic coupling of the Mn moments amounting to $\sim$4.6~$\mu_{\mathrm{B}}$.
The energy difference to the antiferromagnetic configuration is $\sim$14~meV/formula unit.
This magnetic behavior is also reflected in the spin polarized bandstructure in Fig.~\ref{fig4}(\textbf{b}) where both bands at the edges of the gap are derived from majority spin while minority spin states are separated by several hundred meV.
In contrast to the previously reported magnetic non-vdW 2D candidates in Ref.~\cite{Friedrich_NanoLett_2022}, the magnetic Mn ions are not at the surface of the slabs but in the interior (see also Fig.~\ref{fig1}(\textbf{b})).
This is an important difference as the magnetic properties can be expected to be structurally better protected from environmental influences such as adsorbates.
Based on the outlined electronic and magnetic characteristics, these systems can thus reveal potential for \emph{e.g.} optoelectronic and/or spintronic applications.

\noindent
\textbf{{Conclusions}.}
We have outlined a new group of non-vdW 2D materials exhibiting ultra low exfoliation energies --- ultimately getting as small as the one of graphene.
The investigated sulfides Al$_2$S$_3$ and In$_2$S$_3$ have exfoliation energies a factor of four smaller than the corresponding oxides, which can be traced back to exceptionally strong surface relaxations allowing for a significant energy gain.
The smallest values close to the ones of traditional 2D systems are found for SbTlO$_3$ and MnNaCl$_3$, as evident from the comparison to several vdW materials.
The computed band structures of these most easily exfoliable compounds exhibit appealing electronic, possibly topological, and magnetic properties.
Our results may thus be an important guide for extending the family of non-vdW 2D exfoliable systems representing a new class of low dimensional compounds and for studying their characteristics as well as applications.

\section*{Methods} \label{comp_det}

\noindent
The \emph{ab-initio} calculations are performed with \AFLOW\
\cite{curtarolo:art53,aflowPAPER}
and the \underline{V}ienna
\emph{\underline{A}b-initio} \underline{S}i\-mu\-lation
\underline{P}ackage (\VASP) \cite{kresse_vasp,vasp_prb1996,kresse_vasp_1}
employing the exchange-correlation functionals \PBE\ \cite{PBE}, \SCAN\ \cite{Perdew_SCAN_PRL_2015}, \SCAN$+${\small rVV10} \cite{Peng_PRX_2016}, and \PBE$+U$ \cite{Dudarev1998,Liechtenstein1995,Anisimov_Mott_insulators_PRB1991}  with parameter choices in accordance with the \AFLOW\ standard \cite{curtarolo:art104} as well as setting the
internal \VASP precision to {\small ACCURATE}.
For \SCAN,
projector-augmented-wave ({\small{PAW}}) pseudopotentials \cite{kresse_vasp_paw} of \VASP\ version 5.4 are used and non-spherical contributions to the gradient of the density in the {\small{PAW}} spheres are included for \SCAN\ and \PBE$+U$.
The [001]
2D facets are constructed from the
bulk standard conventional unit cell with the respective \AFLOW\ commands \cite{curtarolo:art62} resulting in structures with 10 atoms
and including at least 20~\AA\ of vacuum perpendicular to the slabs.
For all facets, relaxation of both the ionic positions and the cell shape are carried out unless stated otherwise.
The \AFLOW\ internal automatic determination of $k$-point sets is used and for the calculations of the 2D facets, the setting for the number of $k$-points per reciprocal atom \cite{curtarolo:art104} is reduced to 1,000 resulting in $\Gamma$-centered $10\times10\times1$ grids.
The bandstructures are calculated for the optimized \SCAN\ geometry using \PBE($+U$) according to the \AFLOW\ standard \cite{curtarolo:art104} as this functional has been successfully employed previously for the electronic properties of non-vdW 2D systems in Ref.~\cite{Friedrich_NanoLett_2022}.
For computational efficiency, the dynamic stability check through the construction of 2$\times$2 in-plane supercells was carried out with \PBE($+U$).
\par

The bulk and 2D candidate systems with expected magnetic ordering (MnNaCl$_3$, CrGeTe$_3$, CrSiTe$_3$, FePSe$_3$, and MnPSe$_3$) are rigorously checked for magnetism using the algorithm developed within the \CCE\ method \cite{Friedrich_CCE_2019,Friedrich2021}, \emph{i.e.} investigating all possible \FM\ and \AFM\ configurations in the structural unit cell for five different sizes of induced magnetic moments each.
In each case, the lowest energy magnetic state is used for the further calculations.\par

The {exfoliation} energy is computed as:

\begin{equation}
\Delta E_{\mathrm{{exf}}}=\frac{E_{\mathrm{slab}}-E_{\mathrm{bulk}}}{A},
\end{equation}

\noindent
where $E_{\mathrm{slab}}$ and $E_{\mathrm{bulk}}$ indicate the total energies
of the relaxed 2D material and bulk, respectively and $A$ is the {in-plane} surface area according to the relaxed bulk unit cell.
{As proven in Ref.~\cite{Jung_NanoLett_2018}, the exfoliation energy from the surface of the material is exactly equal to the binding energy between layers/facets in the bulk.
}

Numerical data for the {exfoliation} energies are included in
the Supporting Information.

\section*{Acknowledgments}

The authors thank the HZDR Computing Center, HLRS, Stuttgart, Germany and TU Dresden Cluster ``Taurus" for generous grants of CPU time.
R.F. acknowledges support from the Alexander von Humboldt foundation under the Feodor Lynen research fellowship.
A.V.K. thanks the German Research Foundation ({\small DFG}) for the support through Project KR 4866/2-1 and the collaborative research center “Chemistry of Synthetic 2D Materials” SFB-1415-417590517.
R.F. acknowledges Stefano Curtarolo and Agnieszka Kuc for fruitful discussions.
The authors thank Mani Lokamani for technical support.

\section*{Associated content} \label{SI}

\textbf{Supporting Information Available}\\

\noindent
Exfoliation energies for different terminations for ternaries and for comparing results from different functionals, band structures for all binaries and ternaries compared to their bulk parent structures, and tables with numerical data.

\newcommand{\Ozolins}{Ozoli{\c{n}}{\v{s}}}
\providecommand{\latin}[1]{#1}
\makeatletter
\providecommand{\doi}
  {\begingroup\let\do\@makeother\dospecials
  \catcode`\{=1 \catcode`\}=2 \doi@aux}
\providecommand{\doi@aux}[1]{\endgroup\texttt{#1}}
\makeatother
\providecommand*\mcitethebibliography{\thebibliography}
\csname @ifundefined\endcsname{endmcitethebibliography}
  {\let\endmcitethebibliography\endthebibliography}{}

\clearpage


\section*{Supplementary material}

# A New Group of Two-Dimensional Non-van der Waals Materials with Ultra Low Exfoliation Energies
## Supporting Information

Tom Barnowsky,[1,2] Arkady V. Krasheninnikov,[1,3] and Rico Friedrich[1,2,*]

[1]*Institute of Ion Beam Physics and Materials Research,*
*Helmholtz-Zentrum Dresden-Rossendorf, 01328 Dresden, Germany*
[2]*Theoretical Chemistry, Technische Universität Dresden, 01062 Dresden, Germany*
[3]*Department of Applied Physics, Aalto University, Aalto 00076, Finland*
(Dated: September 26, 2022)

## I. Comparing exfoliation energies of ternary systems with different terminations

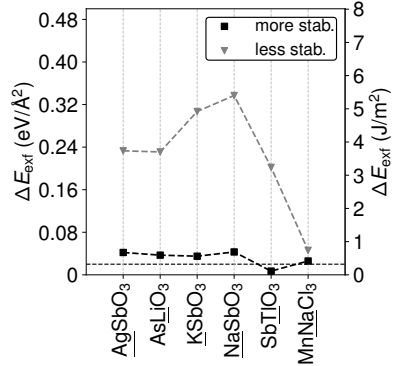

FIG. S1. **Exfoliation energies for different terminations for ternary systems.** The terminating elements for the energetically more stable slabs are underlined. As a reference, the exfoliation energy of graphene [1, 2] is indicated by the dashed black line. The dashed lines connecting the data points are visual guides.

Fig. S1 shows the exfoliation energies for the energetically preferred (more stable) and unfavored (less stable) termination for ternaries. Due to the large difference in the oxidation states of the surface cations for the first five systems ($+1$ vs. $+5$) the values vary by a factor five to almost 30 for the different terminations. For $MnNaCl_3$, the change is less pronounced since due to the $Cl^-$ anions, the inner Mn is assigned the oxidation state $+2$.

## II. Comparing exfoliation energies from different functionals

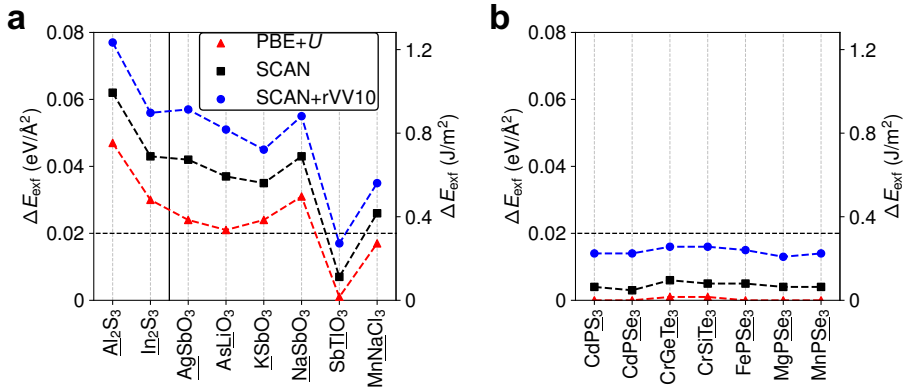

FIG. S2. **Comparison of exfoliation energies.** Exfoliation energies for different functionals for (**a**) the eight non-vdW 2D candidates and (**b**) seven vdW 2D systems with the same structure. Note that in case of $Al_2S_3$, $AsLiO_3$, $KSbO_3$, $NaSbO_3$, and $SbTlO_3$ PBE+$U$ reduces to PBE according to the standard workflow of AFLOW. As a reference, the exfoliation energy of graphene [1, 2] is indicated by the dashed horizontal black line. The dashed lines connecting the data points are visual guides.

## III. Band structures and densities of states

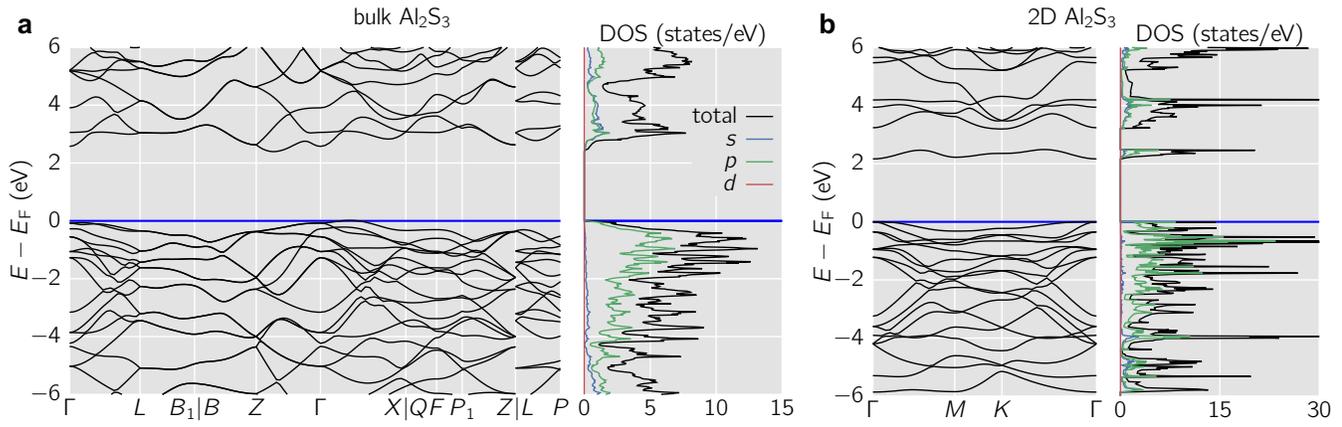

FIG. S3. Bandstructure and density of states for bulk (**a**) and 2D (**b**) $Al_2S_3$. The energies are aligned at the respective Fermi energy $E_F$.

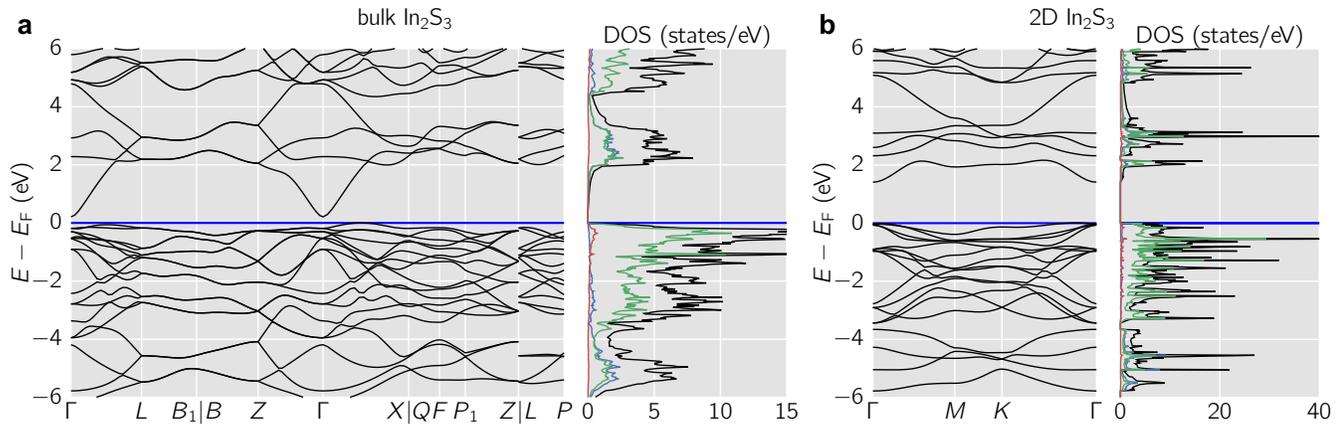

FIG. S4. Bandstructure and density of states for bulk (**a**) and 2D (**b**) $In_2S_3$. The energies are aligned at the respective Fermi energy $E_F$.

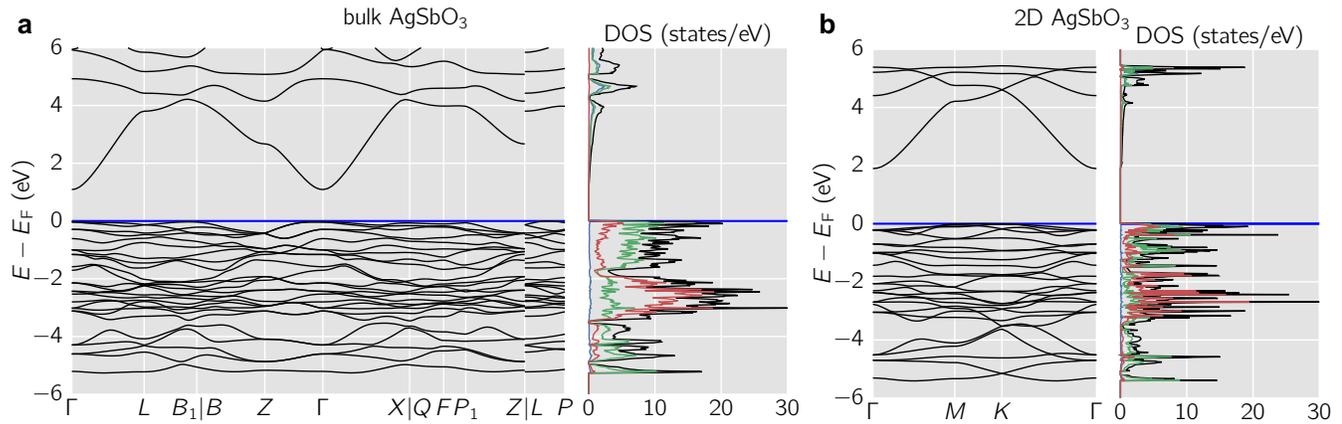

FIG. S5. Bandstructure and density of states for bulk (**a**) and 2D (**b**) $AgSbO_3$. The energies are aligned at the respective Fermi energy $E_F$. For the spin polarized bandstructure, majority spin bands (positive DOS) are indicated in black while minority spin bands (negative DOS) are in red.

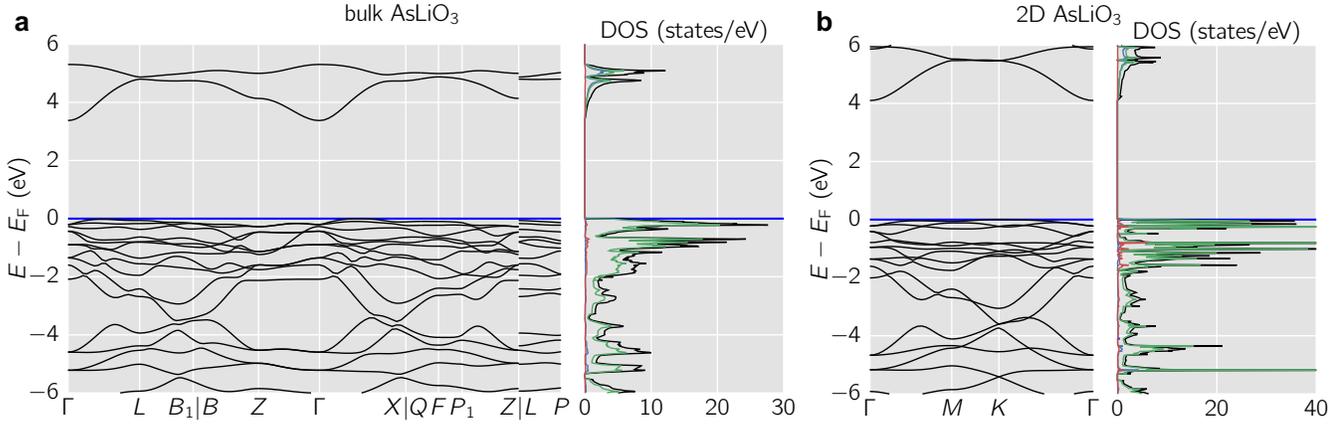

FIG. S6. Bandstructure and density of states for bulk (**a**) and 2D (**b**) AsLiO$_3$. The energies are aligned at the respective Fermi energy $E_F$.

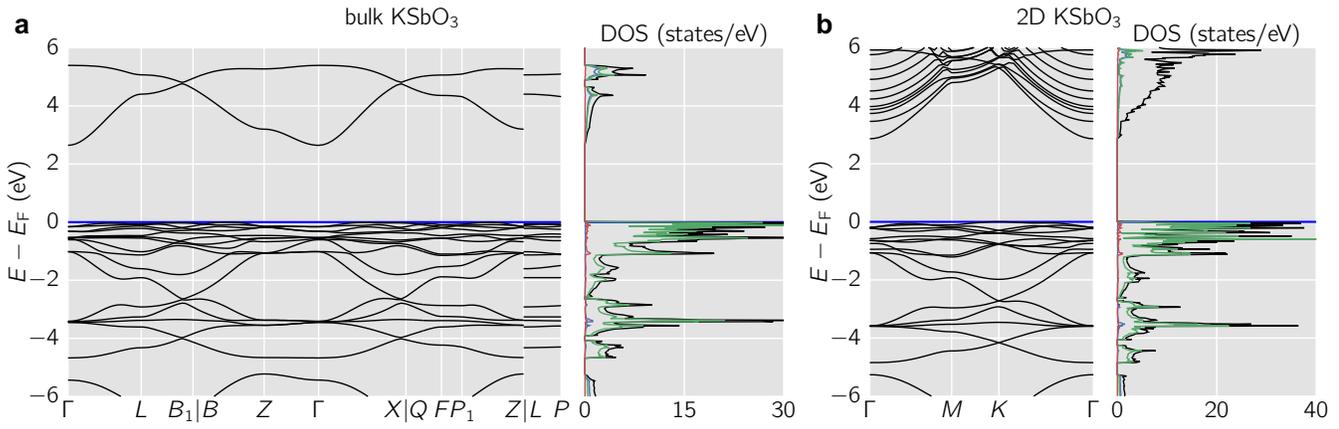

FIG. S7. Bandstructure and density of states for bulk (**a**) and 2D (**b**) KSbO$_3$. The energies are aligned at the respective Fermi energy $E_F$.

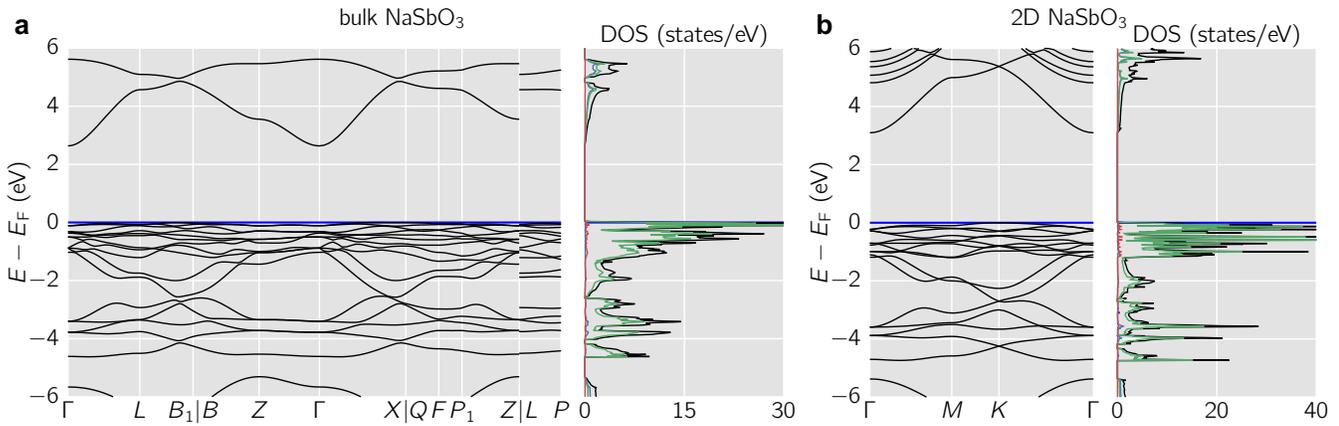

FIG. S8. Bandstructure and density of states for bulk (**a**) and 2D (**b**) NaSb$_3$. The energies are aligned at the respective Fermi energy $E_F$.

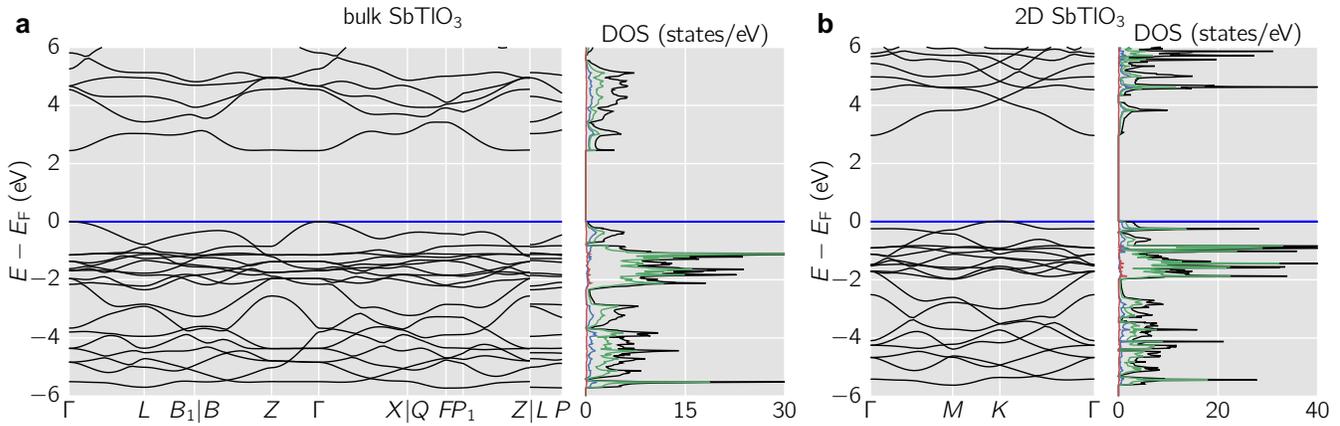

FIG. S9. Bandstructure and density of states for bulk (**a**) and 2D (**b**) SbTlO$_3$. The energies are aligned at the respective Fermi energy $E_\mathrm{F}$.

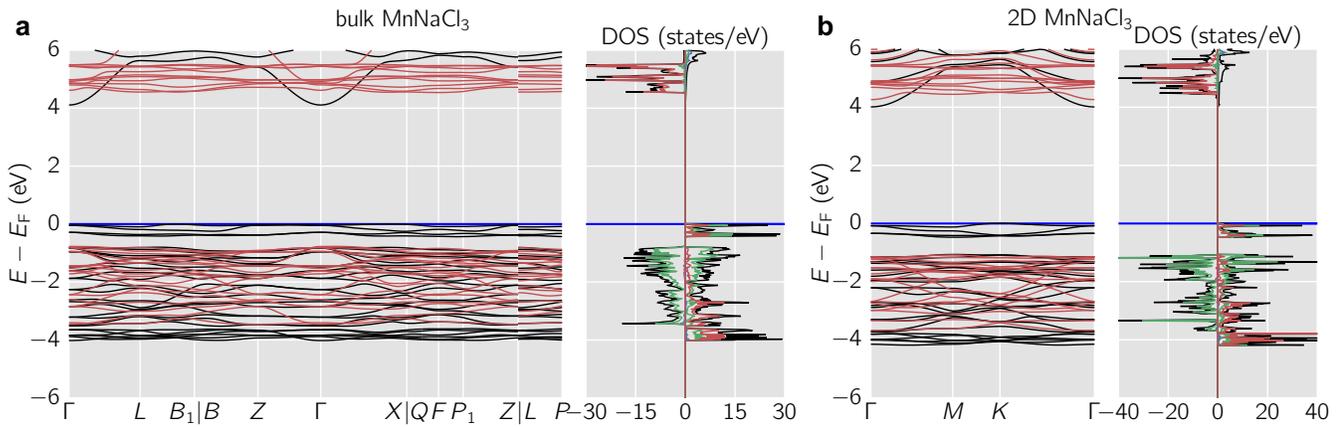

FIG. S10. Bandstructure and density of states for bulk (**a**) and 2D (**b**) MnNaCl$_3$. The energies are aligned at the respective Fermi energy $E_\mathrm{F}$. For the spin polarized bandstructure, majority spin bands (positive DOS) are indicated in black while minority spin bands (negative DOS) are in red.

## IV.  Tables with numerical data

TABLE I: **Exfoliation energies for binaries.** Exfoliation energies for the binaries calculated with different functionals from only a static ("as sliced") electronic calculation as well as when relaxing the ionic positions and (in-plane) cell parameters of the 2D materials. Note that in case of Al$_2$S$_3$ PBE+$U$ reduces to PBE according to the standard workflow of AFLOW. All values are in eV/Å$^2$.

| formula | PBE(+$U$) | | SCAN | | SCAN+rVV10 | |
|---|---|---|---|---|---|---|
| | static | rel. ions and cell | static | rel. ions and cell | static | rel. ions and cell |
| Al$_2$S$_3$ | 0.196 | 0.047 | 0.223 | 0.062 | 0.237 | 0.077 |
| In$_2$S$_3$ | 0.114 | 0.030 | 0.136 | 0.043 | 0.149 | 0.056 |

TABLE II: **Exfoliation energies for ternaries.** Exfoliation energies for the ternary systems calculated with different functionals for fully relaxing all systems, *i.e.* the ionic positions and the (in-plane) cell parameters. The data for the energetically preferred (more stab.) cation termination are given. For SCAN, also the values from only a static ("as sliced") electronic calculation and the energetically less preffered (less stab.) cation termination are included. Note that in case of $AsLiO_3$, $KSbO_3$, $NaSbO_3$, $SbTlO_3$, and $MgPSe_3$ PBE+$U$ reduces to PBE according to the standard workflow of AFLOW. All values are in eV/Å$^2$.

| compound | PBE(+$U$) more stab. | SCAN more stab. static | SCAN more stab. rel. ions and cell | SCAN less stab. rel. ions and cell | SCAN+rVV10 more stab. | compound | PBE(+$U$) | SCAN | SCAN+rVV10 |
|---|---|---|---|---|---|---|---|---|---|
| $AgSbO_3$ | 0.024 | 0.061 | 0.042 | 0.233 | 0.057 | $CdPS_3$ | 0 | 0.004 | 0.014 |
| $AsLiO_3$ | 0.021 | 0.057 | 0.037 | 0.231 | 0.051 | $CdPSe_3$ | 0 | 0.003 | 0.014 |
| $KSbO_3$ | 0.024 | 0.040 | 0.035 | 0.307 | 0.045 | $CrGeTe_3$ | 0.001 | 0.006 | 0.016 |
| $NaSbO_3$ | 0.031 | 0.056 | 0.043 | 0.337 | 0.055 | $CrSiTe_3$ | 0.001 | 0.005 | 0.016 |
| $SbTlO_3$ | 0.001 | 0.007 | 0.007 | 0.202 | 0.017 | $FePSe_3$ | 0 | 0.005 | 0.015 |
| $MnNaCl_3$ | 0.017 | 0.039 | 0.026 | 0.046 | 0.035 | $MgPSe_3$ | 0 | 0.004 | 0.013 |
| | | | | | | $MnPSe_3$ | 0 | 0.004 | 0.014 |


\end{document}